

\message{----------------------------------------------------- }
\message{> > > > M a n u s c r i p t V e r s i o n 2.3 < < < < }
\message{----------------------------------------------------- }



\message{...assuming 8.5 x 11 inch paper}

\magnification=\magstep1	
\raggedbottom
\hsize=6.4 true in
 \hoffset=0.27 true in		
\vsize=8.7 true in

\voffset=0.28 true in         	

\parskip=9pt
\def\singlespace{\baselineskip=12pt}      
\def\sesquispace{\baselineskip=15pt}      




%


%
%
%
%
%
%
%
%
%
%
%
%


%
 \let\miguu=\footnote
 \def\footnote#1#2{{$\,$\parindent=9pt\baselineskip=13pt%
 \miguu{#1}{#2\vskip -7truept}}}
%
%

\def\linebreak{\hfil\break}

\def\Nobreak{\par\nobreak}	

\def\BulletItem #1 {\item{$\bullet$}{#1 }}

\def\AbstractBegins
{
 \singlespace                                        
 \bigskip\leftskip=1.5truecm\rightskip=1.5truecm     
 \centerline{\bf Abstract}
 \smallskip
 \noindent	
 } 
\def\AbstractEnds
{
 \bigskip\leftskip=0truecm\rightskip=0truecm       
 }

\def\section #1 {\bigskip\noindent{\bf #1 }\par\nobreak\smallskip\noindent}

\def\subsection #1 {\medskip\noindent[ {\it #1} ]\par\nobreak\smallskip}

\def\reference{\hangindent=1pc\hangafter=1} 

\def\ref{\reference}

 %

 %

 %

\def\author#1 {\medskip\centerline{\it #1}\smallskip}









%



 at10pt
 %

 %
 %
 %










%

%
%



\def\implies{\Rightarrow}

%


 \def\dal{\displaystyle{{\hbox to 0pt{$\sqcup$\hss}}\sqcap}}



\def\lto{\mathop
        {\hbox{${\lower3.8pt\hbox{$<$}}\atop{\raise0.2pt\hbox{$\sim$}}$}}}
\def\gto{\mathop
        {\hbox{${\lower3.8pt\hbox{$>$}}\atop{\raise0.2pt\hbox{$\sim$}}$}}}
%
%
%



\def\&{{\phantom a}}







\def\interior #1 {  \buildrel\circ\over  #1}     




\def\basisvector#1#2#3{
 \lower6pt\hbox{
  ${\buildrel{\displaystyle #1}\over{\scriptscriptstyle(#2)}}$}^#3}




\def\&{{\phantom a}}		


\def\eeaa{{e^2 a^2\over 6\pi}}



\phantom{}
\vskip -1 true in
\medskip
\rightline{gr-qc/9811030}
\rightline{SU--GP--98/11--1}
\vskip 0.3 true in

\vfill

\bigskip
\bigskip

\sesquispace
\centerline{{\bf An energy bound deduced from the vanishing of}}
\centerline{{\bf the radiation reaction force during uniform acceleration}%
\footnote{$^\star$}{to appear in {\it Modern Physics Letters A}}} 


\singlespace

\bigskip
\centerline {\it Rafael D. Sorkin}
\medskip
 
\centerline {\it Department of Physics, 
                 Syracuse University, 
                 Syracuse, NY 13244-1130, U.S.A.}
\smallskip
\centerline {\it \qquad\qquad internet address: sorkin@physics.syr.edu}

\AbstractBegins 
We deduce from energy conservation a lower bound on the mass of any
system capable of imparting a constant acceleration to a charged body.
We also point out a connection between this bound and the so called
dominant energy condition of general relativity.
\AbstractEnds

\sesquispace
\bigskip\medskip
\noindent
A well known peculiarity of the radiation reaction force on a charged
particle is that (in flat spacetime) it vanishes when the particle
accelerates uniformly.  But this raises a paradox.  An accelerating
charge radiates, and the longer the acceleration continues, the greater
the total energy radiated.  If one asks where this energy comes from in
the case of uniform acceleration, the usual answer is that it is
``borrowed'' from the near field of the particle and then ``paid back''
when the acceleration finally ceases.  But this ``debt'' can be
arbitrarily great if the acceleration remains uniform for a long enough
time.  What, then, if the agent causing the acceleration decides not to
repay the borrowed energy?  What if, in fact, it does not even possess
enough energy to pay its immense debt at that time?  If we believe in
conservation of energy the respective answers must be that the
accelerating agent must not be at liberty to avoid transferring the
required energy and that it must always possess the necessary amount to
cover its accumulated debt.

I claim that this last requirement implies a lower bound on the
mass-energy of the device that impels the accelerating charge forward.
The form of this inequality might be guessed on dimensional grounds
(given that it involves only the mass, the charge, and the magnitude
of the acceleration), but the situation is simple enough that one can
analyze it more exactly.

To make the story more vivid, let us conceive of the accelerating
particle as a billiard ball carrying a fixed surface charge $e$ of
static electricity and of the agency causing the acceleration as a long
rubber band wrapped around the billiard ball (and being tugged on by a
rocket, say).  The acceleration can be very slight, so that our everyday
intuition will be a good guide to what happens.  Suppose that at $t=0$
in some inertial frame, the ball is at rest and we begin to pull the
rubber band with a constant force (in the rest frame of the rubber
band), so that uniform acceleration ensues.  After some years of proper
time, we slip the rubber band off the billiard ball.  It is obvious that
nothing much will happen, and the ball will continue on its way with a
constant velocity equal to the velocity it had at the point of release.
Yet in this process of release, an immense energy must have been
transferred.

Of course the billiard ball itself has acquired an immense kinetic
energy because it is now moving near the speed of light, as is the
rubber band.  However the work done before the release --- given that
the force of radiation reaction has been zero all along --- is by
definition no more than would have been needed to accelerate an
uncharged body of equal mass.  Nothing extra has been provided for the
energy radiated.  Therefore, there is indeed an immense debt.  

At least the debt appears immense when energy is reckoned with respect
to the original reference frame, in which the system began at rest.
When referred to the comoving frame at the moment of release, the debt
might be of a very different magnitude, and in fact one can guess that
it is rather small in that frame.  Nevertheless, it is not zero, and
this means that the rubber band (in which we henceforth include the rest
of the accelerating agency, i.e. the rocket or whatever) must possess
enough energy to transfer the required minimum, willy nilly, to the
system consisting of the billiard ball and its self field.

What is the magnitude of this minimum energy?  For simplicity let the
motion follow the curve $x^2-t^2=1/a^2$ in some system of Galilean
coordinates (so $a$ is the magnitude of the proper acceleration).  The
energy-momentum radiated per unit proper time is known to be
$$
     {dP^a\over d\tau} = \eeaa \, {dx^a\over d\tau}      \eqno(1)
$$
(in units such that $c=\epsilon_0=1$).  Hence the total 4-momentum
radiated from beginning to end is
$$
  P_{radiated}^a =
  \int {dP^a\over d\tau} d\tau 
   = \eeaa \int {dx^a\over d\tau} d\tau
   = \eeaa \Delta{x}^a                   \eqno(2)
$$
where $\Delta{x}^a=(x_{final}-x_{initial})^a$.  On the other hand, if
the final energy transferred is $\Delta{m}$ in the comoving frame (and
if, for example, no momentum is transferred), then the spacetime vector
of transferred four-momentum is
$$
 P_{transferred}^a = \Delta m {dx^a\over d\tau}|_{final}  \eqno(3)
$$

Now, in our coordinate system, the components of the final 4-velocity
$u^a={dx^a/d\tau}$ 
are $(u^0,u^1)=a\,(x,t)$, as is easy to derive from the
facts that $u^a$ is normalized and orthogonal to the radius vector;
and the components of $\Delta x^a$ are 
$(\Delta x^0,\Delta x^1)=(\Delta t,\Delta x)=(t,x-1/a)$.
Furthermore, 
after the particle has been moving at the speed of light for a while, 
we have $t\approx x \gg 1/a$.
Hence
$$
     {dx^a\over d\tau}|_{final} \approx a \,\Delta x^a  \eqno(4)
$$

Comparing equations (2) and (3) in view of
(4), we see that 
$$
    \Delta m \, a \approx \eeaa  
    \quad \implies \quad   
    \Delta m \approx {e^2 a\over 6\pi}
$$ 
In order for this to be possible, 
the rubber band must have at least the mass $\Delta m$, 
so we arrive at the promised lower bound, 
$$
        m  \  \gto \  {e^2 a\over 6\pi}           \eqno(5)
$$
governing the mass-energy of a rubber band impressing an acceleration of
magnitude $a$ on a charge of magnitude $e$.  

In the form given above, the derivation of (5) does little to
suggest an underlying reason for this bound.  However, one can see that
there exists, at least heuristically, a connection between our result
and the energy conditions utilized in general relativity.  We explore
this connection further in the Appendix.


In concluding, I would like to thank Eanna Flannagan and Donald Marolf
for the discussion of uniformly accelerating charges that stimulated
this work.  This research was partly supported by NSF grant PHY-0098488
and by a grant from the Office of Research and Computing of Syracuse
University.


\bigskip\bigskip

\centerline {\bf Appendix}
\Nobreak
Let $F$ be the tension in the rubber band or equivalently the force it
exerts on the billiard ball, and let $\mu$ be its mass per unit length.
The ``dominant energy condition'', requires that $F\le\mu$.  Furthermore,
if $r$ is the radius of the billiard ball, then in order to wrap around
the ball, the rubber band's length should at least be in the vicinity of
$2r$, whence its energy $m$ will be bounded below by $2r\mu$.
Assuming that the energy condition is satisfied, this implies
$m\gto2rF$.\footnote{$^\dagger$}
{One might think to shorten the rubber band by gluing it to the billiard
 ball instead of wrapping it, but in this case the stresses set up in
 the ball --- which we have been neglecting --- might induce a similar
 mass increment.   We have been neglecting any such effect by assuming
 tacitly that the inertial mass of the billiard ball is unaffected by
 acceleration.  If one wished to relax this assumption, then perhaps the
 quantity $m$ in (5) should be re-interpreted as the net mass of
 rubber band plus mass excess induced in the ball.} 

Now let $M$ be the mass of the billiard ball itself.  From the force
equation $F=Ma$ we can conclude $M=F/a{\lto}m/2ra$.  But since the
electric field surrounding the ball contributes an energy of $e^2/8\pi
r$ to $M$, we also have $M\gto e^2/8\pi r$.  Concatenating these
inequalities yields $m \gto e^2a/4\pi$, which is substantially the same
as (5).

%

\end


Outline stuff (put here so doesn't need to be commented out)

(prog1    'now-outlining
  (Outline 
      "
   "\\\\messag" 
     "......"
      "
   "\\\\section."
   "\\\\appendix"
   "\\\\Referenc"	
   "\\\\Abstract" 	
   "\\\\subsection"
   ))